\documentclass[manuscript]{aastex}

\usepackage{amssymb,amsmath}
\usepackage{graphicx}

\shorttitle{Influences of radiation pressures on black hole mass
estimates}

\shortauthors{Liu, Feng, \& Bai}

\begin{document}

\title{Influences of Radiation Pressures on Mass Estimates of Supermassive Black Holes in AGNs}

\author{H. T. Liu\altaffilmark{1,2}, H. C. Feng\altaffilmark{1,3} and J. M. Bai\altaffilmark{1,2}}

\altaffiltext{1}{Yunnan Observatories, Chinese Academy of
Sciences, Kunming, Yunnan 650011, China}

\altaffiltext{2}{Key Laboratory for the Structure and Evolution of
Celestial Objects, Chinese Academy of Sciences, Kunming, Yunnan
650011, China}

\altaffiltext{3}{University of Chinese Academy of Sciences,
Beijing 100049, China}

\email{htliu@ynao.ac.cn}

\begin{abstract}
In this paper, we investigate the influences of two continuum
radiation pressures of the central engines on the black hole mass
estimates for 40 active galactic nuclei (AGNs) with high accretion
rates. The continuum radiation pressure forces, usually believed
negligible or not considered, are from two sources: the free
electron Thomson scattering, and the recombination and
re-ionization of hydrogen ions that continue to absorb ionizing
photons to compensate for the recombination. The masses
counteracted by the two radiation pressures $M_{\rm{RP}}$ depend
sensitively on the percent of ionized hydrogen in the clouds
$\beta$, and are not ignorable compared to the black hole virial
masses $M_{\rm{RM}}$, estimated from the reverberation mapping
method, for these AGNs. As $\beta$ increases, $M_{\rm{RP}}$ also
does. The black hole masses $M_{\rm{\bullet}}$ could be
underestimated at least by a factor of 30--40 percent for some
AGNs accreting around the Eddington limit, regardless of redshifts
of sources $z$. Some AGNs at $z < 0.3$ and quasars at $z \ga 6.0$
have the same behaviors in the plots of $M_{\rm{RP}}$ versus
$M_{\rm{RM}}$. The complete radiation pressures will be added as
AGNs match $M_{\rm{RP}}\ga 0.3 M_{\rm{RM}}$ due to the two
continuum radiation pressures. Compared to $M_{\rm{RM}}$,
$M_{\rm{\bullet}}$ might be extremely underestimated if
considering the complete radiation pressures for the AGNs
accreting around the Eddington limit.

\end{abstract}

\keywords{galaxies: active-- galaxies: nuclei -- galaxies: Seyfert
-- quasars: general -- quasars: supermassive black holes}

\section{INTRODUCTION}
Active galactic nuclei (AGNs), such as quasars and Seyfert
galaxies, are powered by gravitational accretion of matter onto
supermassive black holes in the central engines. The energy
conversion in AGNs is more efficient as implied by high flux
variability and short variability timescales \citep{Ul97}.
Accretion of matter onto black holes can have high energy release
efficiency \citep{Re82,Re84}. Broad-line regions (BLRs) in AGNs
are photoionized by the central radiation of accretion disks. The
broad emission line variations will follow the ionizing continuum
variations due to the photoionization process
\citep[e.g.][]{BM82,Pe93}. The reverberation mapping model was
tested with the reverberation mapping observations
\citep[e.g.][]{Ka99,Ka00,Pe05}. A review about the reverberation
mapping research studies is given by \citet[][and references
therein]{Ga09}. A general assumption for the cloud ensemble is
often virial equilibrium. The BLRs are mainly dominated by the
gravitational potentials of the supermassive black holes. This
would imply Keplerian orbits of clouds, and this was supported
with evidence for the Keplerian motions of the BLR clouds within
the well-studied Seyfert 1 galaxy NGC 5548
\citep[e.g.][]{PW99,Be07}. The virialization assumption has been
commonly and widely accepted for the reverberation mapping
observations to estimate the black hole masses $M_{\rm{\bullet}}$.
This treatment neglects the contribution of a radiation pressure
to the dynamics of clouds.

As the central ionizing source is photoionizing the BLR clouds,
the central radiation will produce a radiation pressure on the
clouds. The radiation pressure will counteract a part of the
gravitational force of a black hole. Thus, the clouds undergo a
decrease of the effective gravity of the black hole, and then the
orbital velocities of the clouds will decrease for the same
orbital radii. In general, this will decrease the widths of broad
emission lines, and then lead to underestimate the black hole
masses based on the reverberation mapping method. The radiation
pressure force may contribute significantly to the cloud motion,
and in this case the clouds on bound orbits could be significantly
sub-Keplerian \citep{Ma08,Ma09,Ne09,NM10,Kr11}. However, most of
the photoionized gas in the BLR follows Keplerian orbits
\citep{Ba14}. \citet{Ma08} studied the effect of radiation
pressure on black hole virial mass estimates for narrow-line
Seyfert 1 galalxies, and the black hole virial masses might be
significantly underestimated if the radiation pressure is
neglected. \citet{Ne09} found that the radiation pressure force is
not important in $0.1\le z \le 0.2$ AGNs with $L(\rm{5100
\rm{\AA}})= 10^{42.8-44.8} \/\ \rm{erg \/\ s^{-1}}$. The simple
virial mass estimates can give a reasonable approximation to
$M_{\rm{\bullet}}$ even when the radiation pressure force is
important \citep{NM10}. In the reverberation mapping researches
\citep[e.g.][]{Ka99,Ka00,Pe05}, the radiation pressures of the
central engines were not considered in estimating
$M_{\rm{\bullet}}$ for the broad-line AGNs. The continuum
radiation pressures due to the free electron Thomson scattering
and the recombination and re-ionization of hydrogen ions are
usually believed negligible or not considered. In this paper, we
will investigate the influences of the continuum radiation
pressures on the black hole mass estimates for AGNs with high
accretion rates, including quasars and broad-line Seyfert 1
galaxies.

The structure of this paper is as follows. Section 2 presents
model. Section 3 presents applications. Section 4 is for
discussion and conclusions. In this work, we assume the standard
$\Lambda$CDM cosmology with $H_0=70 \rm{\/\ km \/\ s^{-1} \/\
Mpc^{-1}}$, $\Omega_{\rm{M}}$ = 0.30, and $\Omega_{\rm{\Lambda}}$=
0.70.

\section{MODEL}
Several assumptions are made. First, the clouds mainly consist of
hydrogen and are partially ionized. Second, the clouds are blobs
at the distance $R$ from the central black hole. Dynamical
equilibrium corresponds to circular orbits \citep[e.g.][]{Kr11}.
Third, the clouds are in the circular orbits. Fourth, the clouds
are optically thin to the Thomson scattering. A cloud has a mass
$m\approx N m_{\rm{H}}$, where $N$ is the total number of hydrogen
in the cloud and $m_{\rm{H}}$ is the mass of hydrogen atom. The
cloud is subject to the gravity of the central black hole with a
mass $M_{\rm{\bullet}}$, and the gravity $F_{\rm{G}}$ is
\begin{equation}
F_{\rm{G}}=\frac{G M_{\rm{\bullet}} m}{R^2},
\end{equation}
where $G$ is the gravitational constant, and $R$ is the orbital
radius of this cloud. The cloud is exerted with a centrifugal
force $F_{\rm{c}}$
\begin{equation}
F_{\rm{c}}=\frac{m v^2}{R},
\end{equation}
where $v$ is the circular velocity. Here, we consider two origins
of continuum radiation pressure forces on a cloud: the Thomson
scattering of the central source radiation by free electrons, and
the recombination and re-ionization of hydrogen ions that will
continue to absorb ionizing photons to compensate for the
recombination. The recombination line photon number per unit time,
i.e., the recombination rate of the ionized hydrogen, is
$v_{\rm{rec}}=n_{\rm{e}}N_{\rm{H^{+}}} \alpha _{\rm{B}}=\beta N
n_{\rm{e}} \alpha _{\rm{B}}$, where $\alpha_{\rm{B}}$ is the
hydrogen recombination coefficient, $n_{\rm{e}}$ is the number
density of the free electrons, $\beta$ is within $0 \leq \beta
\leq 1$, and $N_{\rm{H^{+}}}$ is the number of the hydrogen ions.
On the hydrogen ionizing timescale $\tau_{\rm{ion}}$, the neutral
hydrogen atoms from the recombination are re-ionized into the
hydrogen ions with a number $v_{\rm{rec}} \tau_{\rm{ion}}= \beta N
n_{\rm{e}} \alpha _{\rm{B}}\tau_{\rm{ion}}= \beta N
\tau_{\rm{ion}}/\tau_{\rm{rec}}$, where $n_{\rm{e}} \alpha
_{\rm{B}}=1/\tau_{\rm{rec}}$. Thus, the total continuum radiation
pressure force on the cloud due to the above two origins is
\begin{equation}
F_{\rm{r}}=\frac{L \sigma_{\rm{T}} }{4\pi R^2 c}\beta N+\frac{f L
\sigma_{\rm{bf}} }{4\pi R^2 c}\beta N
\frac{\tau_{\rm{ion}}}{\tau_{\rm{rec}}},
\end{equation}
where $f =L_{\rm{ion}}/L$ is the ratio of the ionizing luminosity
$L_{\rm{ion}}$ to the central source luminosity $L$,
$\sigma_{\rm{T}}$ is the Thomson cross-section, $c$ is the speed
of light, $\beta N$ is the ionized number of hydrogen atoms in the
cloud, $\sigma_{\rm{bf}}$ is the flux-weighted average of the
hydrogen bound-free absorption cross-section, and
$\tau_{\rm{rec}}$ is the hydrogen ion recombination timescale in
the cloud. The two continuum radiation pressure forces in equation
(3) are similar to those in equation (9) in \citet{Kr12} that
focused on stability of BLR clouds. The resultant of forces is
$F_{\rm{t}}=F_{\rm{G}}-F_{\rm{c}}-F_{\rm{r}}$. As $F_{\rm{t}}=0$,
we have
\begin{equation}
\begin{split}
M_{\rm{\bullet}}&=\frac{v^2 R}{G} +\beta \frac{L
\sigma_{\rm{T}}}{4\pi c Gm_{\rm{H}}}+\beta \frac{f L
\sigma_{\rm{bf}}}{4\pi c
Gm_{\rm{H}}}\frac{\tau_{\rm{ion}}}{\tau_{\rm{rec}}}\\
&= \frac{v^2 R}{G}+\frac{L \sigma_{\rm{T}}}{4\pi c Gm_{\rm{H}}}
\left(\beta +
\beta f \frac{\sigma_{\rm{bf}}}{\sigma_{\rm{T}}}\frac{\tau_{\rm{ion}}}{\tau_{\rm{rec}}} \right) \\
&=M_{\rm{BH}}+7.95\times 10^6 \left(\beta + \beta f \frac{
\sigma_{\rm{bf}}}{\sigma_{\rm{T}}}\frac{\tau_{\rm{ion}}}{\tau_{\rm{rec}}}
\right) L_{\rm{45}} \/\
(M_{\rm{\odot}})\\
&=M_{\rm{BH}}+M_{\rm{RP}},
\end{split}
\end{equation}
where $m\approx N m_{\rm{H}}$ is used in the deduction,
$L_{\rm{45}}=L/10^{45} \/\ \rm{erg \/\ s^{-1}}$, $M_{\rm{BH}}$ is
the black hole mass estimated as $L$ is not considered, and
$M_{\rm{RP}}$ is the black hole mass counteracted by the continuum
radiation pressure due to the central engine luminosity.

The cross-section $\sigma_{\rm{bf}}$ is equal to
\begin{equation}
\sigma_{\rm{bf}}= \frac{\int^{\rm{\infty}}_{\rm{\nu_H}}
F_{\rm{\nu}}\sigma_{\rm{\nu}} \rm{d}
\nu}{\int^{\rm{\infty}}_{\rm{\nu_H}} F_{\rm{\nu}} \rm{d} \nu}
=\frac{\int^{\rm{\infty}}_{\rm{\nu_H}}
F_{\rm{\nu}}\sigma_{\rm{\nu}} \rm{d} \nu}{F_{\rm{ion}}}
=\frac{\int^{\rm{\infty}}_{\rm{\nu_H}}
F_{\rm{\nu}}\sigma_{\rm{\nu}} \rm{d} \nu}{fF},
\end{equation}
where $F_{\rm{\nu}}\propto \nu^{-\alpha}$, $\nu_H$ is the
frequency of 13.6 $\rm{eV}$ photon, $\sigma_{\rm{\nu}}$ is the
photoionization cross-section of hydrogen, $F_{\rm{ion}}$ is the
ionizing radiation flux, and $F$ is the total radiation flux.
For $\alpha = 1.5$ \citep{TM73},
$\sigma_{\rm{bf}}\simeq 1.1\times 10^{-18} \/\ \rm{cm^2}$. The
ionizing timescale is
\begin{equation}
\tau_{\rm{ion}}= {\left(\int^{\rm{\infty}}_{\rm{\nu_H}}
\frac{F_{\rm{\nu}}\sigma_{\rm{\nu}}}{h\nu} \rm{d} \nu
\right)}^{-1}.
\end{equation}
The recombination timescale of ionized hydrogen is
\begin{equation}
\tau_{\rm{rec}}= {\left(n_{\rm{e}} \alpha_{\rm{B}}(T)
\right)}^{-1},
\end{equation}
where $n_{\rm{e}}=\beta n_{\rm{H}}$ is the free electron number
density and $n_{\rm{H}}$ is the number density of hydrogen. The
hydrogen recombination coefficient $\alpha_{\rm{B}}(T)$ was given
with a set of formulae in \citet{Se59}. An analytic expression of
$\alpha_{\rm{B}}$ to the excited levels of hydrogen is given by
\citet{HM63}
\begin{equation}
\alpha_{\rm{B}}(T)=1.627\times 10^{-13} T^{-1/2}_4 [1-1.657\log
T_4 + 0.584T^{1/3}_4] \rm{\/\ cm^3 \/\ s^{-1}},
\end{equation}
where $T_4=10^{-4}T$, and $T$ in units of $\rm{K}$ is the electron
temperature of clouds. As $0.5 \le T_4 \le 10$, equation (8) gives
$\alpha_{\rm{B}}$ to an accuracy of better than 1 percent relative
to the result of \citet{Se59}. The photoionized gas in clouds is
at $T\sim 10^4 \/\ \rm{K}$ and produces optical and ultraviolet
emission lines of quasars \citep{Kr81}. Photoionization governs
the thermodynamics of the clouds that have a stable equilibrium
temperature of order $10^4 \/\ \rm{K}$, independent of their
locations \citep{Kr11}. In the case of $T\sim 10^4 \/\ \rm{K}$,
$\alpha_{\rm{B}}\sim 2.6 \times 10^{-13}\rm{\/\ cm^3 \/\ s^{-1}}$.
For the BLRs of AGNs, \citet{Be09a} presented the calibrated
radius-luminosity relation
\begin{equation}
\log R =-21.3+0.519 \log \lambda L_{\rm{\lambda}}(5100 \rm{\AA)},
\end{equation}
where $R$ and $\lambda L_{\rm{\lambda}}(5100 \rm{\AA)}$ are in
units of light days and $\rm{erg \/\ s^{-1}}$, respectively. The
radiation flux $F=L/4\pi R^2$.

In general, there is $0<\beta <1$ that means a partially ionized
cloud. $\beta =1$ means a fully ionized cloud, and $\beta =0$
means a fully non-ionized cloud. First, we consider a case that a
fraction of hydrogen is ionized, and we will take $\beta =0.1$
(Case A). The cloud may be mostly ionized, and we will take $\beta
=0.5$ (Case B). The ratio of $L_{\rm{ion}}/L$ has an average $f
\simeq 0.6$ \citep{Ma08}. A typical value of $n_{\rm{H}} = 10^{10}
\/\ \rm{cm^{-3}}$ was obtained for the BLRs of AGNs
\citep{DN79,FE84,Re89}. The luminosity $L$ is estimated with
$L=9\times \lambda L_{\rm{\lambda}}(5100 \rm{\AA)}$ \citep{Ka00}.
$L=6\times \lambda L_{\rm{\lambda}}(3000 \/\ \rm{\AA})$ is used
for the ultraviolet (UV) continuum \citep[e.g.][]{Wi10}. The range
of $0<\beta <1$ is suggested by the corresponding variations
between the broad-line and continuum light curves in the
reverberation mapping observations \citep[e.g.][]{Ka99,Ka00,Ka05}.
In the following section, we take $f=0.6$, $n_{\rm{H}} = 10^{10}
\/\ \rm{cm^{-3}}$, $T=10^4 \/\ \rm{K}$, and $\alpha = 1.5$ to
estimate $M_{\rm{RP}}$ for Cases A and B.

\section{APPLICATIONS}
We collect 40 high accretion rate AGNs, including quasars and
broad-line Seyfert 1 galaxies, with the black hole virial masses
$M_{\rm{RM}}$ estimated from the reverberation mapping method (see
Table 1). Table 1 consists of four parts. The first part contains
17 AGNs, collected from \citet{Pe04}, with relatively high
accretion rates. The second part contains 13 AGNs with
super-Eddington accreting massive black holes
\citep[SEAMBHs,][]{Du15}. The first results from a new
reverberation mapping campaign are reported in \citet{Du14} for
the SEAMBH AGNs. The new reverberation mapping observations of the
SEAMBH AGNs are presented in \citet{Wa14} and \citet{Hu15}. The
third part contains 9 quasars in the Canada-France High-$z$ Quasar
Survey (CFHQS) at redshift $z\ga$ 6, accreting close to the
Eddington limit \citep{Wi10}. The last one contains 2 AGNs from
other works. There is one overlap AGN (see Table 1). The CFHQS
quasars have $M_{\rm{RM}}$ and luminosity $\lambda
L_{\rm{\lambda}}(3000 \/\ \rm{\AA})$. The optical luminosity
$\lambda L_{\rm{\lambda}}(opt)$ around 5100 $\rm{\AA}$ at the rest
frame of source is used to estimate $L$ with $L=9\times \lambda
L_{\rm{\lambda}}(opt)$ \citep{Ka00}. As in \citet{Wi10},
$L=6\times \lambda L_{\rm{\lambda}}(3000 \/\ \rm{\AA})$ is used
for the CFHQS quasars. The estimated results of $M_{\rm{RP}}$ are
presented in Table 1.

Figure 1 shows the comparisons of $M_{\rm{RP}}$ to $M_{\rm{RM}}$.
For Case A, four AGNs with SEAMBHs and three CFHQS quasars have
$M_{\rm{RP}}\ga 0.5 M_{\rm{RM}}$ (see Figure 1a and Table 1). For
Case A, 9 AGNs match $M_{\rm{RP}}\ga 0.3 M_{\rm{RM}}$ (see Figure
1a). For Case B, all AGNs match $M_{\rm{RP}}\ga 0.5 M_{\rm{RM}}$,
and 30 AGNs follow $M_{\rm{RP}}\ga M_{\rm{RM}}$ (see Figure 1b).
These comparisons show that the continuum radiation pressures of
the central engines have a significant influence on the black hole
virial mass estimates on the basis of the reverberation mapping
method. The masses $M_{\rm{RP}}$ are sensitive to the percent of
ionized hydrogen atoms $\beta$ in the clouds. For some AGNs, such
as IRAS F12397, IRAS 04416, and J1641+3755, the continuum
radiation pressures can counteract about 30--40 percent
gravitational forces of $M_{\rm{\bullet}}$ (Case A).

In the plots of $M_{\rm{RP}}$ versus $M_{\rm{RM}}$, the SEAMBH
AGNs, the CFHQS quasars, and the rest of AGNs populate in the
narrow zones (see Figure 1). From this view point, the SEAMBH AGNs
at $z <0.2$ and the CFHQS quasars at $z\ga 6$ are not essentially
different. This means that their central black holes are likely
accreting at the very high rates, regardless of the environments
surrounding these AGNs, but maybe related to the environments
surrounding the central black holes, such as dusty tori and
accretion disks. Also, some of the rest of AGNs are at the similar
accretion rates to those of the SEAMBH AGNs and the CFHQS quasars.
Whether or not the continuum radiation pressure is considered to
estimate $M_{\rm{\bullet}}$ depends on the ratio of
$M_{\rm{RP}}/M_{\rm{RM}}$. The radiation pressure must be
considered as $M_{\rm{RP}}/M_{\rm{RM}}\ga 0.5$ that will lead to
underestimate $M_{\rm{\bullet}}$ by a factor $\ga 1/3$, i.e.,
$M_{\rm{\bullet}}\ga 1.5 M_{\rm{RM}}$. When
$M_{\rm{RP}}/M_{\rm{RM}}\ga 1.0$, $M_{\rm{\bullet}}$ will be
underestimated by a factor $\ga 1/2$, i.e., $M_{\rm{\bullet}}\ga
2.0 M_{\rm{RM}}$. Thus, the importance of the continuum radiation
pressure can not be neglected in estimating $M_{\rm{\bullet}}$ for
the high accretion rate AGNs.

\begin{figure}[htp]
\begin{center}
\includegraphics[angle=-90,scale=0.4]{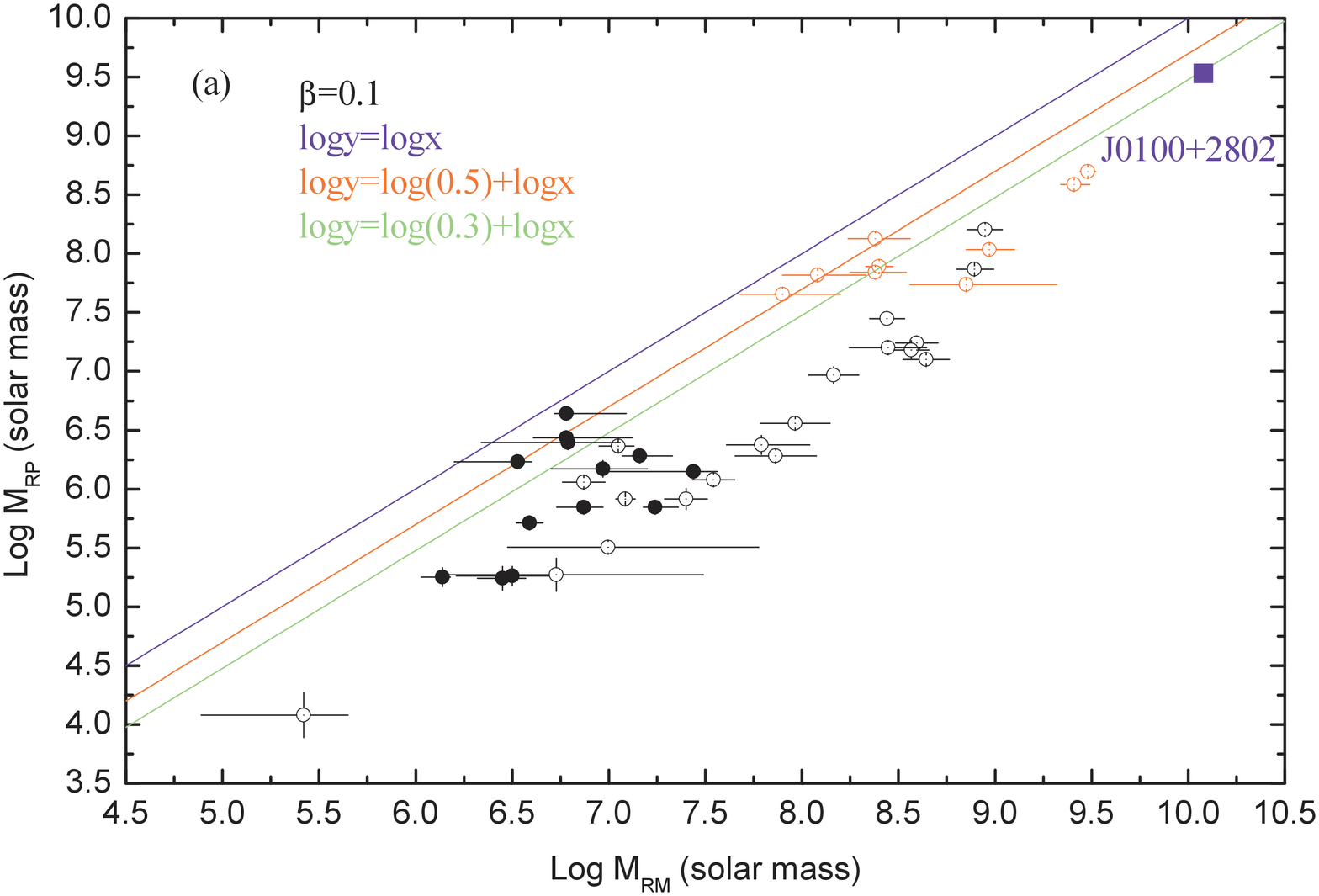}
\includegraphics[angle=-90,scale=0.4]{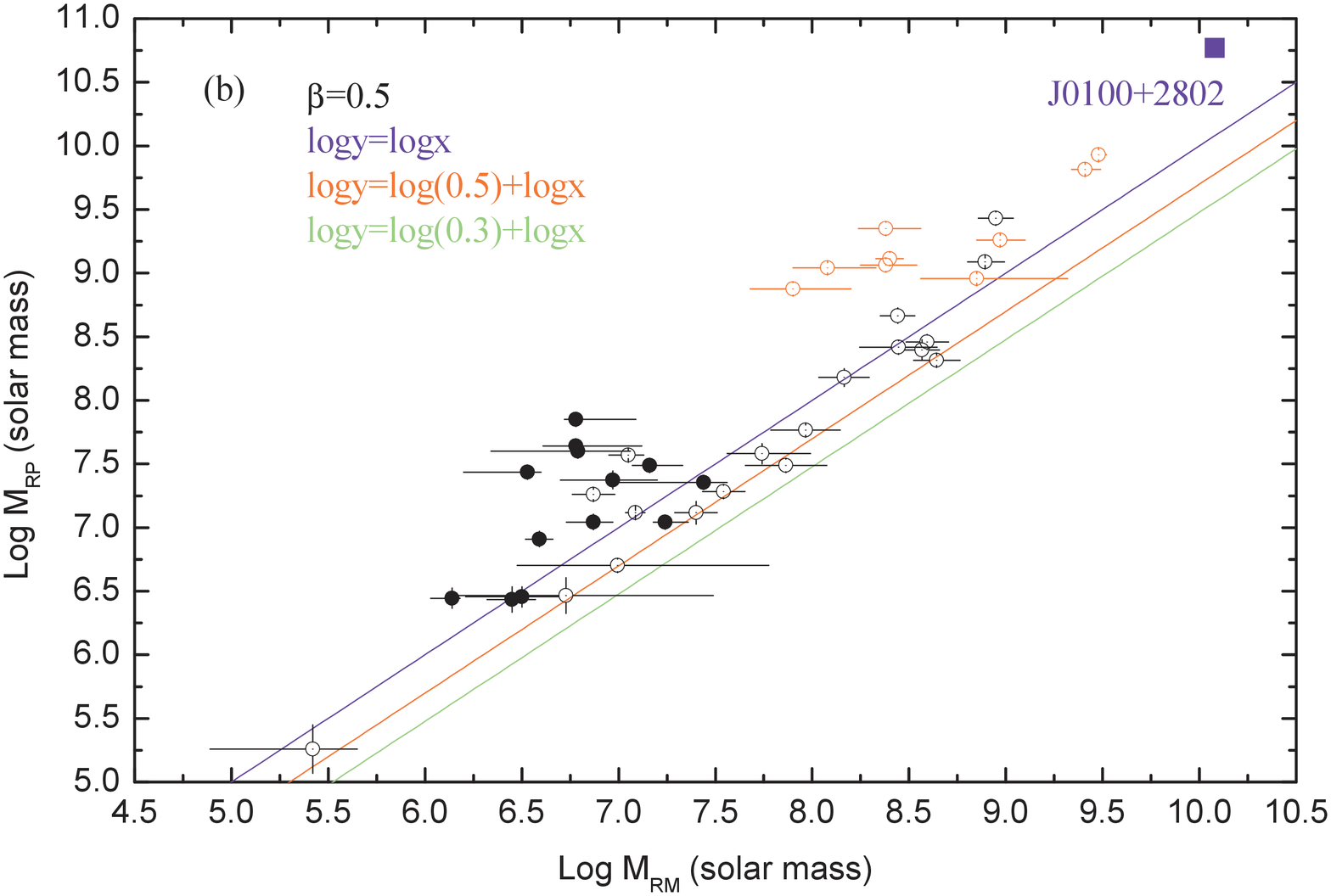}
\end{center}
 \caption{Black hole masses counteracted by radiation pressures versus
 black hole masses estimated with the reverberation mapping method. Solid lines in
 plots (a) and (b) correspond to the different ratios of $y/x$. Solid black circles
 correspond to the AGNs with SEAMBHs \citep{Du15}. Red open circles are
 the CFHQS quasars at $z\sim 6$ \citep{Wi10}.
 Black open circles are the rest of AGNs in Table 1.
 }
  \label{fig1}
\end{figure}

\section{DISCUSSION AND CONCLUSIONS}
There is a undetermined parameter $\beta$, the ionization percent
of the BLR clouds, in equations (3) and (4). Two values of
$\beta=0.1$ and $\beta=0.5$ are assumed to estimate $M_{\rm{RP}}$.
These assumptions may lead to some unreasonable results. Following
the Str\"omgren radius of a Str\"omgren sphere of ionized hydrogen
around a young star \citep{St39}, an ionization depth of a cloud
is defined as the position where the recombination rate equals the
ionization rate, and the hydrogen atoms are fully ionized. As the
recombination and ionization are in equilibrium, the ionizing
photon number arriving at one cloud per unit time equals the
recombination line photon number per unit time. The ionizing
photon number arriving at the cloud per unit time is
$A_{\rm{c}}Q(H)/4 \pi R^2= A_{\rm{c}}U c n_{\rm{H}}$, where
$A_{\rm{c}}$ is the cross-section area of the cloud, $Q(H)=
L_{\rm{ion}}/\langle h\nu \rangle$ is the ionizing photon number
emitted by the central engine per unit time ($\langle h\nu
\rangle$ the mean energy of photons), $U=Q(H)/4 \pi R^2 c
n_{\rm{H}}$ is the ionization parameter of hydrogen, and
$n_{\rm{H}}$ is the hydrogen number density of the cloud. At the
same time, the ionized region of the cloud produces the
recombination line photon number per unit time
$n_{\rm{e}}n_{\rm{H}}\alpha_{\rm{B}}V_{\rm{c,ion}}$, where
$n_{\rm{e}}$ is the electron number density in this region, and
$V_{\rm{c,ion}}$ is the fully ionized volume of the cloud. Thus,
we have $A_{\rm{c}}U c n_{\rm{H}} =
n_{\rm{e}}n_{\rm{H}}\alpha_{\rm{B}}V_{\rm{c,ion}}$ because of the
ionization balance. The ionization depth of the cloud is
$D_{\rm{ion}}\simeq V_{\rm{c,ion}}/A_{\rm{c}}=
Uc/n_{\rm{e}}\alpha_{\rm{B}}= Uc/n_{\rm{H}}\alpha_{\rm{B}}$. The
cloud has $\beta = V_{\rm{c,ion}}/V_{\rm{c}}\simeq
D_{\rm{ion}}/D_{\rm{c}}$, where $V_{\rm{c}}$ and $D_{\rm{c}}$ are
the volume and thickness of the cloud, respectively.

The ionization parameter has $U\sim 0.1$--1 for emission line gas
in quasars and Seyfert galaxies \citep{Da72,Mc75,Kw81}. A gas
density of $10^9$--$10^{10} \/\ \rm{cm^{-3}}$ is required by a
very large Lyman continuum optical depth \citep[e.g.][]{Kr81}. The
gas density values are the same as those in \citet{Kw81} for
quasars and Seyfert galaxies. \citet{FE84} got
$n_{\rm{H}}=10^{10\pm 1} \/\ \rm{cm^{-3}}$. A density of $10^9 \la
n_{\rm{H}}\la 10^{11} \/\ \rm{cm^{-3}}$ is set by the presence of
broad semi-forbidden line C~{\sc III}$]\lambda 1909$ and the
absence of broad forbidden lines such as [O~{\sc III}$]\lambda
\lambda 4363,4959,5007$ \citep{DN79,Re89}. The typical values are
$n_{\rm{H}}\sim 10^{10} \/\ \rm{cm^{-3}}$ and $U\sim 0.1$
\citep[e.g.][]{Re89,La06}. The cloud size of $r_{\rm{c}}\la
(1$--$3)\times 10^{12} \/\ \rm{cm}$ was constrained by the
smoothness of the emission-line profiles \citep{La06}.
\citet{Ba14} showed a universal ionization parameter $U\sim 0.1$
in the inner photoionized layer of the BLR clouds, independent of
luminosity and distance. The radiation pressure confinement of the
photoionized layer appears to explain the universality of the BLR
properties in AGNs, the similar relative line strength over the
vast range of $10^{39}$--$10^{47}$ $\rm{erg \/\ s^{-1}}$ in
luminosity \citep{Ba14}. A widely accepted power-law relation has
been established between the luminosity $\lambda
L_{\rm{\lambda}}(opt)$ and the BLR size $R$, as $R \propto \lambda
L_{\rm{\lambda}}^{0.5}(opt)$
\citep{Be06,Be09a,Be09b,De10,Ka96,Ka00,Ka05,Ka07,Gr10,WZ03}. This
power-law relation spans over a range of $10^7$ in $\lambda
L_{\rm{\lambda}}(opt)$ \citep{Ka07}. Thus, $U = L_{\rm{ion}}/4\pi
R^2 c n_{\rm{H}} \langle h\nu \rangle \propto \lambda
L_{\rm{\lambda}}(opt)/R^2  n_{\rm{H}}$. So, $U$ will be
independent of $\lambda L_{\rm{\lambda}}(opt)$, $L_{\rm{ion}}$,
and $R$. This independence may lead to a universal ionization
parameter as suggested in \citet{Ba14}.

The typical values of $n_{\rm{H}}\sim 10^{10} \/\ \rm{cm^{-3}}$
and $U\sim 0.1$ are taken to estimate $D_{\rm{ion}}$, and then
$\beta$ with $\alpha_{\rm{B}}\sim 2.6 \times 10^{-13}\rm{\/\ cm^3
\/\ s^{-1}}$ and $D_{\rm{c}}\sim 2 r_{\rm{c}}\la (2$--$6) \times
10^{12} \/\ \rm{cm}$. We have $D_{\rm{ion}}\sim 1.2\times 10^{12}$
$\rm{cm}$, and then $\beta \simeq D_{\rm{ion}}/D_{\rm{c}}\ga $
0.2--0.6. Thus, it is basically reasonable to take $\beta =0.1$
and 0.5 in the calculations of $M_{\rm{RP}}$. So, the masses
counteracted by the continuum radiation pressures of the central
engines are not negligible compared to, or are comparable to the
black hole virial masses at least for some AGNs (see Figure 1).
This counteracting effect of the continuum radiation pressure is
significant for quasars at $z\ga 6$ (see Figure 1). J0100+2802 at
$z=6.30$ has a bolometric luminosity of $L=1.62 \times 10^{48} \/\
\rm{erg \/\ s^{-1}}$ and a black hole mass of $\sim 1.2\times
10^{10} \/\ M_{\odot}$, and it is the most luminous quasar known
at $z> 6$ \citep{Wu15}. It has a $M_{\rm{RP}}\sim 1.1 \times
10^{9.5} \/\ M_{\odot}$ as $\beta =0.1$ and $M_{\rm{RP}}\sim 5.9
\times 10^{10} \/\ M_{\odot}$ as $\beta =0.5$. These $M_{\rm{RP}}$
for this object are comparable to the black hole virial mass $\sim
1.2\times 10^{10} \/\ M_{\odot}$. So, J0100+2802 will have
$M_{\rm{\bullet}}> 1.2\times 10^{10}M_{\rm{\odot}}$. This larger
black hole mass further gives rise to the most significant
challenge to the Eddington limit growth of black holes in the
early Universe \citep{Vo12,Wi10}. There are the same cases for the
CFHQS quasars as in J0100+2802 (see Figure 1). Their masses
$M_{\rm{\bullet}}$ are larger than the virial masses
$M_{\rm{RM}}$. \citet{Wa10} suggested for $z\simeq 6$ quasars that
the supermassive black holes in the early Universe likely grew
much more quickly than their host galaxies. The larger black hole
masses $M_{\rm{\bullet}}$ of the quasars at $z\ga 6$ further
strengthen this suggestion. Thus, it is important to consider the
radiation pressure effect on the black hole mass estimates for the
AGNs with the high accretion rates. This importance of the
radiation pressure is consistent with that suggested by
\citet{Ma08} who argued for narrow-line Seyfert 1 galalxies, and
seems to be inconsistent with the suggestions in \citet{Ne09} and
\citet{NM10}.

An assumption of isotropy in equations (3) and (4) is made for the
central source luminosity $L$. This assumption might be
significantly influence the estimated values of $M_{\rm{RP}}$, and
then the results of this paper. The anisotropic illumination of a
BLR by the central radiation was discussed by \citet{Ne87}, and
recently was investigated by \citet{WQ14}. The central radiation
of quasars are anisotropic \citep[e.g.][]{NB10}. \citet{NB10}
derived bolometric corrections based on the theoretical accretion
disk models of \citet{Hu00}, and obtained an anisotropic
correction factor of $f_{\rm{ani}} \approx 0.8$, where
$f_{\rm{ani}}=L/L_{\rm{iso}}$ and $L_{\rm{iso}}$ is the total
source luminosity assuming isotropy. \citet{Ru12} got a factor of
$f_{\rm{ani}} \approx 0.75$ for quasars. So, we take $f_{\rm{ani}}
= 0.8$, i.e., $L=0.8 L_{\rm{iso}}$ to re-estimate $M_{\rm{RP}}$,
and the re-estimated $M_{\rm{RP}}$ are compared to $M_{\rm{RM}}$
in Figure 2. Considering the anisotropy of the central source
luminosity $L$, the continuum radiation pressure effects on the
black hole masses $M_{\rm{\bullet}}$ can not be still neglected
for the close- and super-Eddington limit accretion rate AGNs.
Thus, the anisotropy of the central source luminosity could not
influence significantly the main results assuming isotropy.
\citet{Wa14} proposed the SEAMBH AGNs as the most luminous
standard candles in the Universe. Since a part of the SEAMBH AGNs
and the CFHQS quasars have the same behaviors in the plots of
$M_{\rm{RP}}$ versus $M_{\rm{RM}}$, the CFHQS quasars likely have
the potential to be the most luminous standard candles in the
Universe. This potential will extend the redshifts of the most
luminous standard candles from $z< 0.2$ up to $z\ga 6$, and will
be important to study the Universe because of extending
significantly the cosmic distance beyond the range explored by
type Ia supernovae \citep{Ri98,Pe99}.
\begin{figure}[htp]
\begin{center}
\includegraphics[angle=-90,scale=0.4]{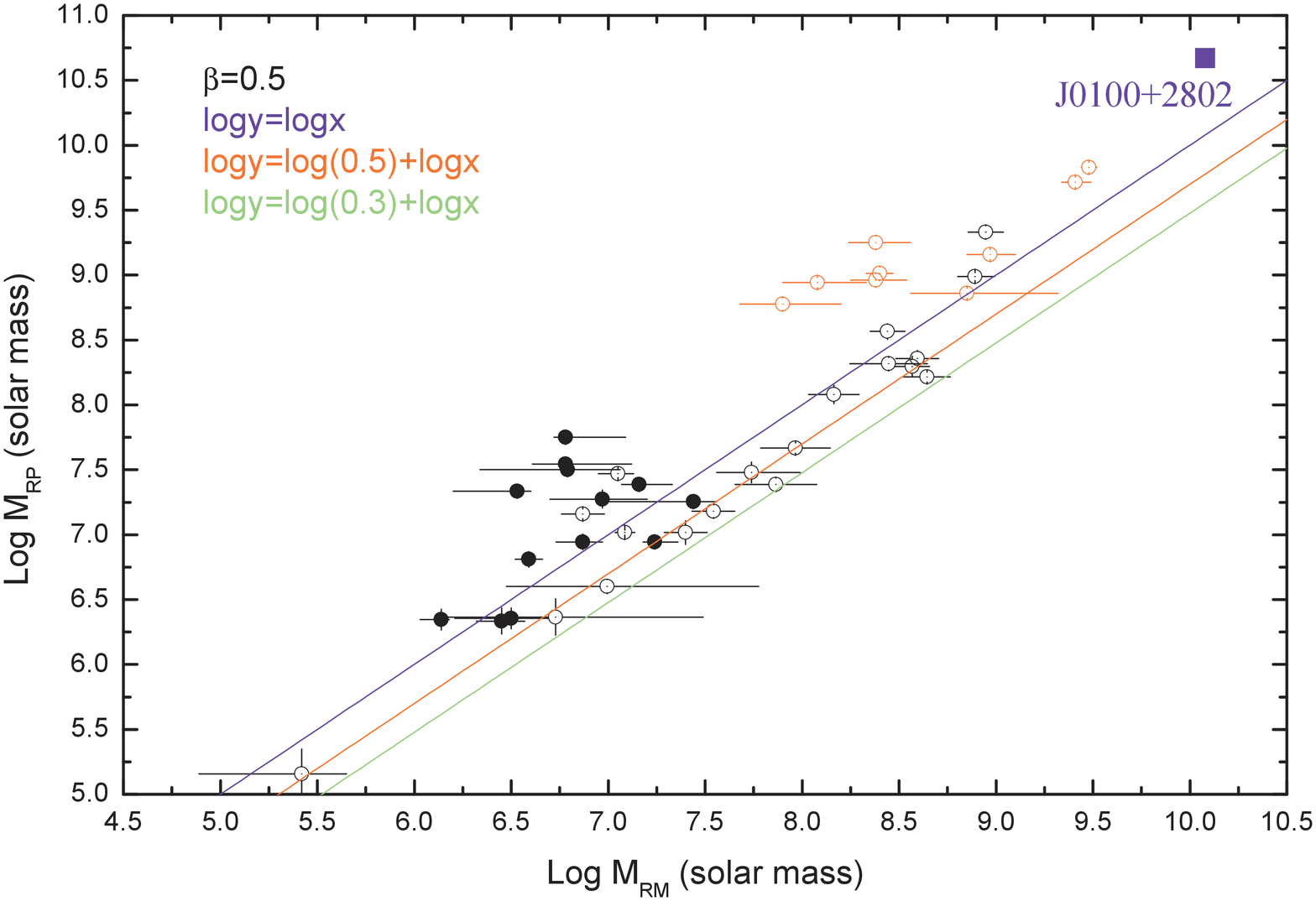}
\end{center}
 \caption{$M_{\rm{RP}}$ versus $M_{\rm{RM}}$ for the same sources as in Figure 1.
 The symbols and lines are the same as in Figure 1. $M_{\rm{RP}}$ are estimated
 as considering the anisotropy of the central engine luminosity $L$.
 }
  \label{fig1}
\end{figure}

An average column density of a cloud is $N_{\rm{H}}\sim 10^{22}
\/\ \rm{cm^{-2}}$ ($n_{\rm{H}}\sim 10^{10} \/\ \rm{cm^{-3}}$ and
$D_{\rm{c}}\sim (2$--$6)\times 10^{12} \/\ \rm{cm}$). The electron
Thomson scattering optical depth of the cloud is
$N_{\rm{H}}\sigma_{\rm{T}} \sim 10^{-2} \ll 1$, and the cloud is
optically thin, which confirms the fourth assumption of the model.
The ionizing optical depth of the cloud is
$N_{\rm{H}}\sigma_{\rm{bf}} \sim 10^{4} \gg 1$, and the cloud is
optically thick to the ionizing photons. The cloud has the
Str\"omgren radius smaller than its thickness. So, the cloud is
partially ionized, confirming the first assumption of the model,
and this is consistent with the constraints set by the
reverberation mapping observations. The radiation pressure force
ratio of the second to the first term in equation (3) has averages
$\simeq 1.2$--6.3 for AGNs in Table 1. The first term of the free
electron Thomson scattering is comparable to the second term of
the recombination and re-ionization of the hydrogen ions. The
second term becomes more important as $\beta$ increases. The
partly ionized clouds will be ionized more as $L_{\rm{ion}}$
increases, and the second term would be more important as the
central engines become brighter. The radiation pressure forces due
to absorption of ionizing photons and Thomson scattering were used
to modify the expression for the black hole virial mass
$M_{\rm{RM}}$ \citep[see equation 5 in][]{Ma08}. As
$N_{\rm{H}}\sim 10^{22}$--$10^{23} \/\ \rm{cm^{-2}}$, the modified
$M_{\rm{RM}}$ becomes larger by a factor of $\sim 10$--100 for the
AGNs accreting around the Eddington limit. In consequence,
J0100+2802 at $z=6.30$ will have $M_{\rm{\bullet}}$ $\sim
10^{11}$--$10^{12} \/\ M_{\odot}$ that is nearly impossible to the
Eddington limit growth of black holes in the early Universe
because of the larger initial masses of primary black holes. If
considering the line-driven radiation pressure force \citep{Ca75},
the radiation pressure force due to the gas opacity will be $\sim
10^3$ times that due to the electron scattering opacity
\citep{Fe09}. This will bring a larger mass $M_{\rm{\bullet}}\sim$
$M_{\rm{RP}}$ $\sim 10^{12}$--$10^{13} \/\ M_{\odot}$ for
J0100+2802. These extremely large corrected masses of the black
hole in J0100+2802 indicate some problem of their treatments of
the radiation pressure forces on the BLR clouds in AGNs, or the
model of formation and evolution of black hole, or the model of
the Universe.

In this paper, we investigate the influences of two continuum
radiation pressures, usually believed negligible or not
considered, of the central engines in AGNs on the black hole mass
estimates with the reverberation mapping method or the descendent
methods. The continuum radiation pressure forces are from two
origins: the free electron Thomson scattering of the central
radiation, and the recombination and re-ionization of the ionized
hydrogen. The radiation pressures depend on a parameter $\beta$,
the ionized percent of the clouds in a BLR. The counteracted black
hole masses by the radiation pressures $M_{\rm{RP}}$ are compared
to the black hole virial masses $M_{\rm{RM}}$ for 40 AGNs with the
high accretion rates. The masses $M_{\rm{RP}}$ are sensitive to
$\beta$. As $\beta =0.5$, $M_{\rm{RP}}\ga 0.3 M_{\rm{RM}}$ for all
the AGNs in Figure 2. As $\beta =0.1$, the radiation pressures can
counteract at least about 30--40 percent gravitational forces of
the black holes for some AGNs (see Figure 1). Four SEAMBH AGNs at
$z < 0.2$ and five CFHQS quasars at $z\ga 6.0$ are around the line
$M_{\rm{RP}}=0.5 M_{\rm{RM}}$ for $\beta =0.1$ (see Figure 1a).
Thus, the continuum radiation pressures of the central engines
have to be considered in estimating the black hole masses for the
AGNs accreting around the Eddington limit, regardless of the
redshifts or the surrounding environments of AGNs. The most
luminous quasar J0100+2802 likely has the same case as the nine
AGNs (see Figures 1 and 2). A part of the SEAMBH AGNs and the
CFHQS quasars is blended with the non-SEAMBH and non-CFHQS AGNs
(see Figures 1 and 2). The close- and super-Eddington limit
accreting AGNs are not different from the rest of AGNs. The
anisotropy of the central source luminosity could not influence
significantly the main results assuming isotropy. The force
multiplier, the ratio of gas opacity to electron scattering
opacity \citep{Ca75,Fe09}, will be needed for the AGNs with
$M_{\rm{RP}}\ga 0.3 M_{\rm{RM}}$ due to the radiation pressures in
equation (3). Though, some extremely large masses are derived from
the force multiplier $\sim 10^3$ for the black hole in J0100+2802
($M_{\rm{RP}}\sim 10^{12}$--$10^{13} \/\ M_{\odot}$). In future,
the force multiplier, due to the photoionization absorption, and
the resonance and subordinate line absorption, could be calculated
with the spectral simulation code Cloudy described by
\citet{Fe98}, and with detailed parameters of the BLR structure,
the cloud distribution, the gas density of cloud, the chemical
abundances of gas, the central continuum, and the observed
emission lines, for each high accretion rate AGNs.

\acknowledgements H.T.L. thanks the National Natural Science
Foundation of China (NSFC; grants 11273052 and U1431228) for
financial support. J.M.B. acknowledges the support of the NSFC
(grant 11133006). H.T.L. thanks the financial supports of the
Youth Innovation Promotion Association, CAS and the project of the
Training Programme for the Talents of West Light Foundation, CAS.

\clearpage

\begin{deluxetable}{rrrrrrrr}

\tablecolumns{7}

\tabletypesize{\scriptsize}


\tablecaption{Sample of AGNs with the virial black hole masses
\label{tbl-1}}

\tablewidth{0pt}

\tablehead{ \colhead{Name} & \colhead{$z$} & \colhead{$\log \frac{\lambda L_{\rm{\lambda}}(opt)}{\rm{erg \/\ s^{-1}}}$} &
\colhead{$\log \frac{M_{\rm{RM}}}{\rm{M_{\rm{\odot}}}}$} & \colhead{Refs.} &\colhead{$\log \frac{M_{\rm{RP}}}{\rm{M_{\rm{\odot}}}}$}&
\colhead{$\log \frac{M_{\rm{RP}}}{\rm{M_{\rm{\odot}}}}$}
\\

\colhead{(1)}&\colhead{(2)}&\colhead{(3)}&\colhead{(4)}&\colhead{(5)}&\colhead{(6)}& \colhead{(7)} }

\startdata

Mrk 335    &0.026&$43.86\pm 0.04$ &$7.15\pm 0.11$& 1  &$6.06\pm0.04$  &$7.26\pm0.04$  \\

PG 0026+129&0.142& $45.02\pm 0.06$& $8.59\pm0.11$ & 1  & $7.24\pm0.06$ &$8.46\pm0.06$ \\

PG 0052+251&0.155& $44.96\pm 0.08$&$8.57\pm0.09$& 1  & $7.18\pm0.08$ &$8.40\pm0.08$  \\

3C 120     &0.033&$44.17\pm  0.08$&$7.74^{+0.25}_{-0.18}$&  1  &$6.37\pm0.08$ &$7.58\pm0.08$  \\

PG 0844+349&0.064&$44.35\pm 0.04$&$7.97\pm 0.18$& 1 & $6.56\pm0.04$ &$7.77\pm0.04$  \\

Mrk 110  &0.035&$43.72\pm 0.09$&$7.40\pm 0.11$& 1 & $5.92\pm0.09$ &$7.12\pm0.09$   \\

PG 0953+414&0.234&$45.22\pm 0.06$&$8.44\pm 0.09$&  1  & $7.45\pm0.06$ &$8.67\pm0.06$  \\

PG 1211+143&0.081&$44.75\pm 0.07$&$8.16\pm 0.13$& 1  & $6.97\pm0.07$ &$8.18\pm0.07$   \\

PG 1226+023&0.158&$45.96\pm 0.05$&$8.95\pm 0.09$& 1  & $8.20\pm0.05$ &$9.43\pm0.05$    \\

PG 1229+204&0.063&$44.08\pm 0.05$&$7.86\pm 0.21$& 1  &$6.28\pm0.05$ &$7.49\pm0.05$ \\

NGC 4593& 0.009& $43.09\pm  0.14$&$6.73^{+0.76}_{-0.56}$&  1&$5.27\pm0.14$ &$6.47\pm0.14$  \\

PG 1307+085&0.155&$44.88\pm 0.04$&$8.64\pm 0.12$&  1  &$7.10\pm0.04$ &$8.31\pm0.04$  \\

IC 4329A& 0.016&$43.32\pm 0.05$&$7.00^{+0.78}_{-0.52}$&  1  &$5.51\pm0.05$&$6.70\pm0.05$  \\

Mrk 279 & 0.030&$43.88\pm 0.05$&$7.54\pm 0.11$& 1   & $6.08\pm0.05$&$7.28\pm0.05$  \\

PG 1613+658&0.129 &$44.98\pm 0.05$&$8.45\pm 0.20$&  1  & $7.20\pm0.05$ &$8.42\pm0.05$   \\

PG 1700+518&0.292&$45.63\pm 0.03$&$8.89^{+0.10}_{-0.09}$&  1   &$7.87\pm0.03$ &$9.09\pm0.03$  \\

NGC 7469 &0.016 &$43.72\pm  0.02$&$7.09\pm 0.05$&   1 & $5.92\pm0.02$ &$7.12\pm0.02$ \\

\hline

$*$Mrk 335    &0.026&$43.65\pm 0.06$&$6.87^{+0.10}_{-0.14}$&  2&$5.84\pm0.06$  &$7.04\pm0.06$ \\

Mrk 1044 &0.017 & $43.06\pm 0.10$&$6.45^{+0.12}_{-0.13}$&    2&$5.24\pm0.10$&$6.43\pm0.10$ \\

Mrk 382  & 0.034&$43.08\pm 0.08$&$6.50^{+0.19}_{-0.29} $&   2&$5.26\pm0.08$&$6.46\pm0.08$   \\

Mrk 142 & 0.045&$43.52\pm 0.06$&$6.59^{+0.07}_{-0.07} $&   2&$5.71\pm 0.06$&$6.91\pm 0.06$  \\

\bf IRAS F12397&0.043&$44.19 \pm 0.05$& $6.79^{+0.27}_{-0.45}$&  2&$6.40\pm  0.05$&$7.60\pm  0.05$  \\

Mrk 486&0.039&$ 43.65\pm 0.05$&$7.24^{+0.12}_{-0.06}$&   2  &$5.84\pm 0.05$&$7.04\pm 0.05$   \\

Mrk 493&0.031&$ 43.07\pm 0.08$&$6.14^{+0.04}_{-0.11}$&   2&$5.25\pm 0.08$&$6.45\pm 0.08$  \\

\bf IRAS 04416&0.089&$  44.43\pm 0.03$& $6.78^{+0.31}_{-0.06}$&  2&$6.64\pm 0.03$&$7.85\pm 0.03$  \\

SDSS J075101&0.121&$44.08 \pm 0.05$&$7.16^{+0.17}_{-0.09}$&   2&$6.28\pm 0.05$&$7.49\pm 0.05$    \\

\bf SDSS J080101&0.140&$44.23\pm 0.03$&$6.78^{+0.34}_{-0.17}$&  2&$6.44\pm 0.03$&$7.64\pm 0.03$  \\

SDSS J081441&0.163&$43.97 \pm 0.07$&$6.97^{+0.23}_{-0.27}$&  2&$6.17\pm 0.07$&$7.37\pm 0.07$  \\

SDSS J081456&0.120&$43.95 \pm 0.04$&$7.44^{+0.12}_{-0.49}$&  2&$6.15\pm 0.04$&$7.35\pm 0.04$   \\

\bf SDSS J093922&0.186&$44.03\pm 0.04$&$6.53^{+0.07}_{-0.33}$&  2&$6.23\pm 0.04$&$7.44\pm 0.04$  \\

\hline

\bf J0210-0456 & 6.438 & $45.60\pm 0.05$$^{\dag}$ &$7.90^{+0.30}_{-0.22}$ & 3 & $7.66\pm 0.05$& $8.88\pm 0.05$ \\

J2329-0301  & 6.417 & $45.83\pm 0.05$$^{\dag}$ &$8.40^{+0.07}_{-0.07}$ & 3 & $7.89\pm 0.05$& $9.11\pm 0.05$ \\

J0050+3445  & 6.253 & $46.51\pm 0.04$$^{\dag}$&$9.41^{+0.08}_{-0.07}$  & 3 & $8.59\pm 0.04$& $9.82\pm 0.04$\\

J0221-0802  & 6.161 & $45.68\pm 0.04$$^{\dag}$&$8.85^{+0.47}_{-0.29}$ & 3 & $7.74\pm 0.04$& $8.96\pm 0.04$ \\

\bf J2229+1457  & 6.152 & $45.76\pm 0.04$$^{\dag}$&$8.08^{+0.25}_{-0.18}$ & 3 &$7.82\pm 0.04$& $9.04\pm 0.04$ \\

J1509-1749  & 6.121 & $46.62\pm 0.04$$^{\dag}$ &$9.48^{+0.04}_{-0.04}$& 3 & $8.70\pm 0.04$& $9.93\pm 0.04$ \\

J2100-1715  & 6.087 & $45.97\pm 0.04$$^{\dag}$ &$8.97^{+0.13}_{-0.12}$ & 3& $8.03\pm 0.04$& $9.26\pm 0.04$ \\

\bf J1641+3755  & 6.047 & $46.06\pm 0.05$$^{\dag}$ &$8.38^{+0.18}_{-0.14}$& 3 & $8.13\pm 0.05$& $9.35\pm 0.05$ \\

J0055+0146 & 5.983 & $45.78\pm 0.05$$^{\dag}$&$8.38^{+0.16}_{-0.13}$ & 3 &$7.84\pm 0.05$ & $9.06\pm 0.05$\\

\hline



NGC 4051 &0.002&$41.92\pm 0.19$&$5.42^{+0.23}_{-0.53}$&  2,4,5&$4.08\pm 0.19$&$5.26\pm 0.19$  \\





PG 2130+099&0.063 &$44.16\pm 0.03$&$7.05^{+0.08}_{-0.10} $&  2,4,6&$6.36\pm 0.03$&$7.57\pm 0.03$   \\

\enddata

\tablecomments{Column 1: AGN names; Column 2: redshifts of
objects; Column 3: Optical luminosity mainly around 5100
$\rm{\AA}$ at rest frame; Column 4: the black hole masses
estimated from the reverberation mapping observations; Column 5:
the references for columns 3 and 4; Column 6: the black hole
masses counteracted by the continuum radiation pressure as $\beta
=0.1$; Column 7: the black hole masses counteracted by the
radiation pressure as $\beta =0.5$; The sign $*$ denotes the other
measured results for the same source as in the first part of Table
1.
The sign $^{\dag}$ denotes the UV luminosity around 3000 $\rm{\AA}$ at the rest frame of source. \\
\textbf{References}: (1) \citealt{Pe04}, (2) \citealt{Du15}, (3)
\citealt{Wi10}, (4) \citealt{Be13}, (5) \citealt{De10}, (6)
\citealt{Gr12}.}

%

\end{deluxetable}

\clearpage

\end{document}